\documentstyle[twocolumn,epsfig]{mn}

\def\ltsima{$\; \buildrel < \over \sim \;$}
\def\lsim{\lower.5ex\hbox{\ltsima}}
\def\gtsima{$\; \buildrel > \over \sim \;$}
\def\gsim{\lower.5ex\hbox{\gtsima}}

\title[Modelling Self-Interacting CDM Haloes with a  
       Cosmological Boltzmann Code]
{ Modelling Self-Interacting CDM Haloes with a  
       Cosmological Boltzmann Code}

\author[Firmani, D'Onghia \& Chincarini] 
{Firmani C.$^{1}$, D'Onghia E.$^{2}$ \& Chincarini G.$^{3,4}$\\ 
$^{1}$ Instituto de Astronom{\'\i}a, UNAM, A.P. 70-264, 04510
M\'{e}xico D.F., M\'{e}xico \\
$^{2}$ Universit\'{a} degli Studi di Milano, via Celoria 16,
 20100 Milano, Italy\\
$^{3}$ Osservatorio Astronomico di Brera,
 via E. Bianchi 46, 23807 Merate (LC), Italy\\
$^{4}$ Universit\'{a} degli Studi di Milano-Bicocca, Italy\\
E--mail: {\tt firmani@astroscu.unam.mx}, {\tt elena@merate.mi.astro.it},
{\tt guido@merate.mi.astro.it}
}

\date{\underline{submitted to MNRAS on September 18, 2000}}

\begin{document}
\maketitle

\begin{abstract}
We investigate the density profiles and evolution of weakly
self-interacting cold dark matter haloes using a numerical
code based on the collisional Boltzmann equation. This approach is
alternative to N-body techniques in 
following the dynamical evolution of haloes in the cosmological context and 
taking into account
particle self-interaction. The physical case with a 
cross section inversely proportional to the
relative velocity of the colliding particles is modelled
with an unprecedented resolution, spanning five orders 
of magnitude on the radius for each halo.
The modelled haloes cover a mass range from 
dwarf galaxies to galaxy clusters. We find that for 
$\sigma  v_{100} \approx 10^{-24}$ cm$^2$/GeV, where $\sigma$
is the cross section per unit mass and $v_{100}$ is the collision 
velocity in units of 100 km/s, soft cores in good agreement 
with observations on galactic as well as on galaxy cluster scales
are obtained. Remarkably, the observed nearly invariance of the halo
central density with mass is reproduced. 

\end{abstract}
\begin{keywords}

cosmology:dark matter - galaxies:formation - galaxies:haloes, 
  methods:kinetic theory
\end{keywords}

\section{Introduction}
The current Cosmology based on the growth of small
fluctuations through gravitational instability in a universe
with cold dark matter (CDM) defines a physically well-motivated theory,
capable to explain most of the properties of the large-scale
structures of the universe. 
In this scenario,
where a large fraction of the matter is supposed non-dissipative, 
cold and collisionless, the dark haloes form hierarchically
via gravitational collapse of primordial density fluctuations.
The density profiles of these dark haloes have
been studied systematically using 
N-body simulations (Navarro, Frenk $\&$ 1996, 1997; hereafter NFW). 
It was found that a simple analytical density profile
with one free parameter is able to describe the typical 
structure of the virialized CDM haloes. On its own, this free 
parameter, which can be the concentration, depends only on the halo mass.

The NFW density profile goes as $\rho \propto r^{-1}$ at small radii.
According to recent high-resolution simulations, the inner 
density profile of the haloes is even steeper than
$r^{-1}$ ($\rho \propto r^{-1.5}$) (Moore et al. 1999; Jing \& Suto
2000). This prediction of the CDM scenario is in potential conflict
with observations: the halo inner density profiles inferred from
the HI rotation curves of dwarf and low surface brightness (LSB) galaxies
and from a mass map constructed by strong lensing techniques for 
the galaxy cluster CL0024+1654 seem to be much shallower than 
$r^{-1}$ (Moore 1994; Flores \& Primack 1995; Burkert 1995; de Blok 
\& McGaugh 1997; Tyson et al. 1998). 
Recently, the rotation curves of 
some of the dwarf and LSB galaxies were re-obtained using
high-resolution $H\alpha$ observations, resulting the inner density 
profiles more concentrated than those obtained from the HI 
observations, but on average still shallower than $r^{-1}$ and 
with a large scatter (Swaters et al. 2000; 
van den Bosch \& Swaters 2000; Dalcanton \& Bernstein 2000). 
Besides, as Firmani et al. 2000b have shown, the halo inner density
profiles of most of the observed LSB galaxies could be shallower
than reported if the baryon contraction is taken into account
(see also Firmani \& Avila-Reese 2000).

An interesting property to be confirmed with more observational
studies, in particular for galaxy clusters, is that the central 
density of dark matter haloes is roughly independent of the halo 
mass with a value around 0.02 M$_{\odot}$ pc$^{-3}$ 
(Firmani et al. 2000a,b). Not only the predictions of the CDM
scenario, but also of alternative scenarios as the warm dark
matter one are unable to predict such a scale invariance of the
central density of dark haloes (Avila-Reese et al. 1998; 
 Col\'{\i}n et al. 2000).

In a recent burst of activity, several alternatives to the 
well motivated CDM scenario were proposed in order to overcome
the difficulties mentioned above and others as the excess of 
satellites and a little 
angular momentum in the galaxy disc predicted by numerical
simulations. These
{\it alternative} theories try to suppress the excess of central mass 
of the galactic haloes and may be summarized into two classes:
1) a manipulation of the power spectrum with a suppression of 
 the small scales as an intrinsic property of a dark matter:
 Avila-Reese et al. 1998; Moore et al. 1999; Firmani et al. 2000a,b;
 Peebles 2000; Col\'{\i}n et al. 2000.
These works make different assumptions concerning thermal energy
at the beginning of the mass aggregation history (MAH).
2) Assuming for dark matter particles a different nature from collisionless:
  self-interacting (Spergel $\&$ Steinhardt 2000); 
repulsive (Goodman 2000); decaying (Cen 2000); with a limiting 
phase-space 
density (Hogan $\&$ Dalcanton 2000) and annihilating (Kaplinghat 
 et al. 2000). Intriguingly, only the assumption
 of a self-interacting or annihilating cold dark matter seems to be able
 to reproduce the observed central density  scale invariance ranging from
 galaxy to galaxy cluster size.

In a previous work, adopting the 
self-interacting CDM scenario suggested by
Spergel \& Steinhardt, we have  proposed a 
thermodynamical model able to describe the gravothermal 
expansion of the core in the haloes (Firmani et al. 2000a); 
 with this model, the
scale invariance of the halo central densities was predicted. 
The NFW density profile shows a velocity dispersion raising outwards
with radius (Cole $\&$ Lacey 1996; Fukushige $\&$ Makino 1997). 
Therefore, if particles are self-interacting,
the inward heat transfer induces a thermalization process able
to avoid the formation of a cuspy core. The influence of 
{\it weak} self-interaction on the structure of the CDM haloes
has been recently explored using cosmological 
numerical simulations (Yoshida et al. 2000; Dav\'e et al. 2000).
These works, where  
constant cross sections were used, have led
to interesting results, that we will discuss in comparison 
with ours, later.

A cross section inversely proportional to the particle
velocity dispersion seems to be better motivated by observations
(Firmani et al. 2000a,b; Wyithe et al. 2000). 
In this
 paper we focus on detailed dynamical predictions of the density
 profiles of self-interacting spherical symmetric 
 dark matter haloes ranging from 
galaxy to galaxy cluster mass.
The cosmological model we assume is
flat with vacuum density $\Omega_{\Lambda}=0.7$ and expansion rate
$H_0=65$  km  s$^{-1}$ Mpc$^{-1}$. 

\section{The cosmological Boltzmann code}

Our dynamical approach aimed to investigate the cosmological 
halo density profiles is based on a solution
 of the collisional Boltzmann 
equation for dark matter. The 
virialized dark halo is described by the distribution function $N(E,J)$, where 
$N(E,J) \Delta E  \Delta J$ gives the number of particles 
within the intervals $\Delta E$ and $\Delta J$  
centred at the total energy E and the angular momentum J. The particle 
population is assumed to be uniform with respect to the directions of {\bf J}. 
The 
mass aggregation supplies new particles, each one with a total energy and 
angular momentum. The computation of each particle orbit is based on a 
gravitational field derived from the mass distribution in a self-consistent
 way. The 
change in time of the gravitational field due to the mass aggregation modifies 
the particle orbit in agreement with its adiabatic invariants. The  
particle mass is distributed along its orbit according to the fraction 
of the period 
spent in any interval. For collisionless particles this approach is  
similar to the {\it secondary infall} (Avila-Reese et al. 1998) and 
allows to calculate a virialized halo 
density profile for mass and angular momentum aggregation histories in a 
specific CDM universe. For a given universe, the mass aggregation 
history (MAH) is obtained making use of the extended Press-Schechter formalism 
based on the conditional probabilities for a Gaussian random field. 
For a given 
mass, we generate a set of MAHs through Monte Carlo simulations and we 
calculate the average MAH. If the perihelion to aphelion ratio of the orbits 
(angular momentum) is opportunely selected, the NFW profile is obtained on 
galaxy and galaxy cluster scales.

The effects of the collisions between particles is described by the 
collisional term of the Boltzmann 
equation and induce a further evolution on the distribution function. Our 
scheme will provide an accurate solution only when the collision time is 
sufficiently larger than the orbital period of the particles. 
The collision cross 
section multiplied the relative velocity of the colliding particles is assumed 
 to be constant in agreement with the previous discussion 
in Firmani et al. 2000b. 
This hypothesis 
makes particularly efficient the  Monte Carlo method implemented to describe 
the collisions of the particles of each discrete state (E,J). 
By each trial we move a 
large amount of particles from one state to another according to the probability 
of the process. The results show an enormous space resolution obtained with a 
minimum computational expense, with each run taking approximately
a couple of hours on a 450MHz pentium processor.

The efficiency of our method is 
counterbalanced by the practical difficulty to explore 
solutions with collision 
times lesser than the particle orbital periods 
(high cross section), as well as 
halo structures which deviate from the spherical symmetry. For these cases 
complementary studies with N-body techniques have been planned.

\section{Self-interacting dark matter halo evolution}

We have calculated the evolution of haloes in a $\Lambda$CDM cosmology
assuming a self-interaction cross section per unit mass ($\sigma$) given 
by: $\sigma \ v_{100} =10^{-24}$ cm$^2$/GeV, where $v_{100}$ is the 
relative velocity of the colliding particles in units of  100 km s$^{-1}$.
Fig. 1 shows the present-day density profiles of haloes with masses
of 6 \ 10$^{11}$ and $6 \ 10^{15}$ M$_{\odot}$ (solid lines). For
comparison, the corresponding NFW profiles are also shown (dotted lines).
Remarkably, the self-interacting CDM haloes appear to be 
thermalized in the centre resulting in a non-singular
isothermal sphere, while in the outer parts, the density
profile is well described by the NFW shape which is representative 
of the hierarchical merger history. The non-singular isothermal 
profile obtained with our code confirms the
 halo profile (dashed line) we derived  in a previous 
work in order to describe observations (Firmani et al. 2000b).
In that case, the hydrostatic equilibrium was assumed in 
order to pass from a NFW profile for the outer regions to an isothermal 
sphere in the centre.
 For that ``naive'' model the core-radius was obtained
once assigned the central density. 
\begin{figure}
\epsfig{file=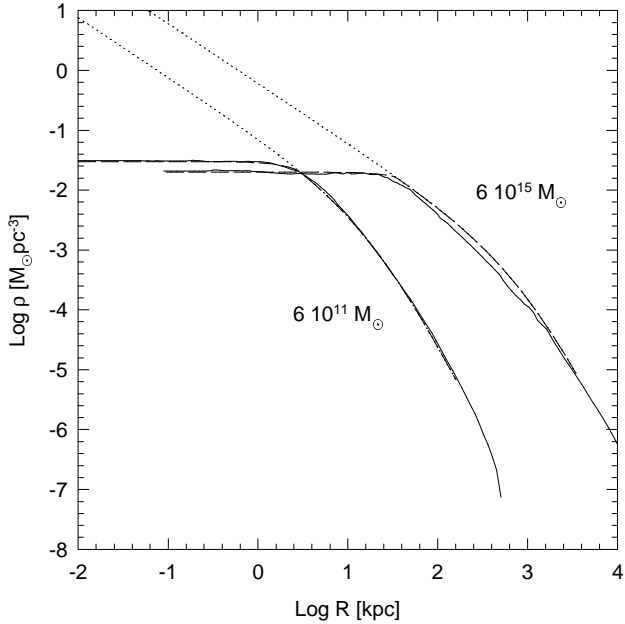,angle=0,width=\hsize,bbllx=67pt,bblly=155pt,
bburx=492pt, bbury=587pt, clip=} 
\@ 
\caption{The behaviour of the mass density profile 
  for haloes with total mass: $M=6 \ 10^{11}$ and $6 \ 10^{15} \
  M_{\odot}$ (solid lines) with the self-interacting 
  cross section: $\sigma \ v_{100}=10^{-24}$ cm$^2$ GeV$^{-1}$.
  Dotted lines show the NFW profile for both haloes. Long-dashed lines 
  draw the non-singular isothermal profile obtained in 
  Firmani et al. 2000b (see the text)}
\end{figure}

With the dynamical model based on the solution of the transfer 
equation we presented here, we can now estimate the typical number 
of collisions in the halo core over a Hubble time ($t_H$): 
$N_{coll}=\rho \sigma v t_H$. For a 6 \ 10$^{11}$ M$_{\odot}$ halo
and the assumed cross section, we find $N_{\rm coll} \approx 4$.

Fig. 2 shows the evolution of the density profile of an halo of
6 \ 10$^{11}$M$_{\odot}$. Since early epochs
($z \approx 40$) a soft core of few pc size is formed. 
At $z=20$ the central 
density is about $4 \ M_{\odot}$ pc$^{-3}$ and the core size is 
already of $\sim 40$ pc. The central density is high enough to allow 
collisions on time scales of only a few hundred million years.
Due to the mass aggregation, the inner radial 
velocity dispersion gradient is preserved. 
Then, due to heat transfers inwards, the soft core
continues growing while the central density decreases. At any time, 
the core density is determined by a self-regulating mechanism which makes
the collision efficiency able to transfer enough
energy to expand the core. The central density decreases and the
core size rises, both reaching values comparable with the present
observations, i.e., $0.03 \ M_{\odot}$ pc$^{-3}$ and 6 kpc,
respectively.   
\begin{figure}
\epsfig{file=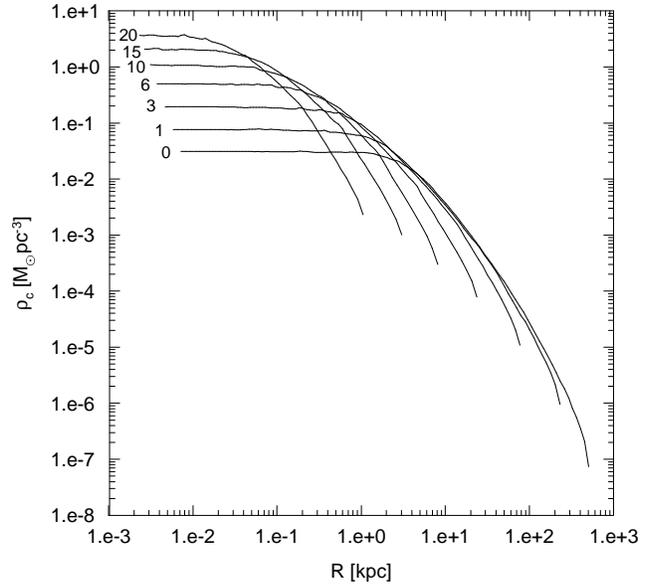,angle=0,width=\hsize,bbllx=57pt,bblly=160pt,
bburx=511pt, bbury=595pt, clip=} 
\@ 
\caption{The evolution in redshift of the density 
 profile is shown for an halo with $M=6 \ 10^{11} \ M_{\odot}$.}
\end{figure}

Of great interest is the invariance of the central density,
shown in Fig.1,      
in passing from an halo of galactic size to an halo with
mass four order of magnitude larger, representative of
a cluster of galaxies.
In Fig. 3a and Fig.3b we compare the central densities and core radii 
predicted by the models with those inferred from 
observations of dwarfs (filled squares), LSB galaxies (empty squares) and 
the cluster CL0024+1654 (filled circle) (see Firmani et al. 
2000b for details). Big empty squares are for those LSB galaxies with
$H\alpha$ rotation curves (Swaters et al. 2000). The models were 
calculated for two cross sections: $\sigma v_{100}=10^{-24}$ and 
$2 \ 10^{-24}$ cm$^2$ GeV$^{-1}$ (dashed and dotted lines, respectively).
\begin{figure}
\epsfig{file=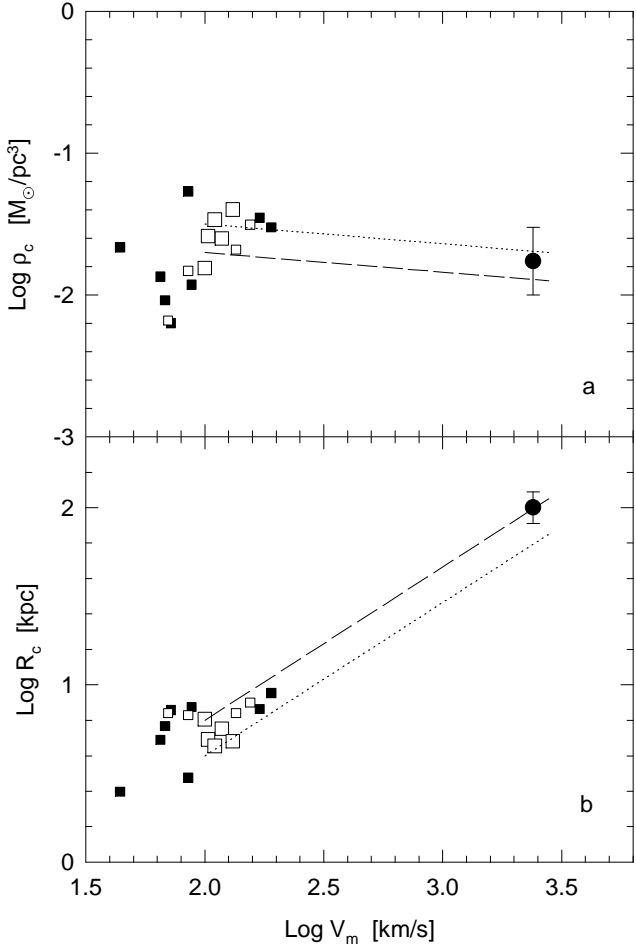,angle=0,width=\hsize,bbllx=95pt,bblly=126pt,
bburx=472pt, bbury=692pt, clip=}
\@
\caption{The plot shows a comparison between central densities 
 and core-radii predicted by the model and the same quantities 
 inferred by available observations of dwarfs (filled squares),
 LSB galaxies (empty squares) and the cluster CL0024+1654 (filled
 circle) (see Firmani et al. 2000b for details). Big empty squares 
 represent LSBs $H\alpha$ rotation curves (Swaters et al. 2000).
 Models are shown for $\sigma \ v_{100}=10^{-24}$ (dotted lines) 
 and $2 \ 10^{-24}$ cm$^2$ GeV$^{-1}$ (dashed lines).}
\end{figure}
\begin{figure}
\epsfig{file=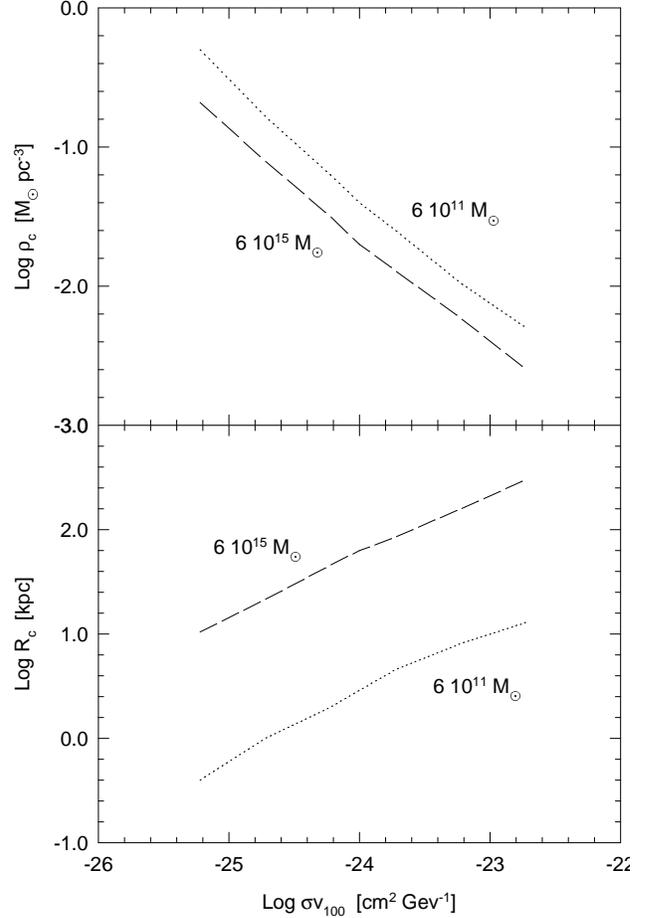,angle=0,width=\hsize,bbllx=54pt,bblly=53pt,
bburx=493pt, bbury=698pt, clip=}
\@
\caption{The top panel shows the central densities of virialized
haloes of: $6 \ 10^{11} M_{\odot}$ (dotted-line) and $6 \ 10^{15}
M_{\odot}$ (dashed-line) as a function of the strenght of the self-interacting
cross section.
In the bottom panel core radii are plotted for the same haloes.}
\end{figure}

The agreement of the self-interacting CDM model with observations
is rather remarkable. In a range of scales from dwarfs to galaxy
clusters, the haloes have (1) a nearly constant central density,
and (2) a core radius roughly proportional to the maximum
circular velocity. We stress that the only way to obtain halo 
central densities independent of their masses is assuming
$\sigma \propto 1/v$. 
 
It is interesting to note that constraining the 
density for the cluster to the value of $\rho_c=0.02 \ M_{\odot}$ pc$^{-3}$
as inferred by the observation of CL0024+1654, we derive 
an estimate of the cross section:  
$\sigma v_{100} \approx  10^{-24}$ cm$^2$/GeV.

\section{Comparison with previous works}

In Fig.4 we show a grid of models for two different masses:
$6 \ 10^{11} M_{\odot}$ and $6 \ 10^{15} M_{\odot}$ and 
wide range of cross sections.
In order to scale the models with the mass for a given value 
of $\sigma v_{100}$, Fig.3  shows that $\rho_c$
is roughly scale invariant while $R_c$ is proportional to $V_m$.

With the aim to compare our models, derived with the assumption
$\sigma \propto v^{-1}$ to models with $\sigma=const$, we will
assume the latter velocity dispersion as representative of the collisional
velocity. 
From Fig.2 of Yoshida and co-workers (2000), extrapolating 
the profile towards the center for the model
S1W-a, with the same pattern of S1W-b and S1W-c, we read
a central density of $\rho_c=2-3 \ 10^{16} M_{\odot}$/h$^2$/Mpc$^3$ 
and a core radius $R_c=35-50$ kpc/h for a halo of 
$7.4 \ 10^{14} \  M_{\odot}$/h.
We have simulated a cluster
of $M=6 \ 10^{15} \  M_{\odot}$. Rescaling the mass of the 
Yoshida et al. (2000) halo ($7.4 \ 10^{14} M_{\odot}$/h and $v=1500$ km/s) 
to the cluster  mass used here 
($6 \ 10^{15} M_{\odot}$ and $v=2400$ km/s), it is easy to obtain   
$\rho_c=0.013-0.008  M_{\odot}$/pc$^3$ and $R_c=85-120$ kpc.
In our model, for the 
cross section of S1W-a, $\sigma=0.1$  cm$^2$/g, we find
$\rho_c=0.01  M_{\odot}$/pc$^3$ and  $R_c=100$ kpc.
Let us now analyze the case 
S1W-b of Yoshida et al. (2000). In this model, which resolution 
is higher than the S1W-a case, the central density is 
$\rho_c=4 \ 10^{15} M_{\odot}$h$^2$ 
/Mpc$^3$ and the core radius $R_c=150$ kpc/h.
Making the same mass rescaling we obtain 
$\rho_c=1.7 \ 10^{-3}  M_{\odot}$/pc$^3$ and  $R_c=370$ kpc.
This case is for a cross section $\sigma=1$  cm$^2$/g. In 
Fig.4 we see that our model for the same $\sigma$ predicts 
$\rho_c=2 \ 10^{-3} M_{\odot}$/pc$^3$ and  $R_c=355$ kpc.
Taking into account the different approaches 
the agreement is really outstanding. 

Regarding Dav\'e et al (2000) work, the only case we have 
simulated is the first top panel from left to right of
Fig.1 of the same paper. For $\sigma=10^{-24}$ cm$^2$ GeV$^{-1}$ 
and for an halo
of $M=6.17 \ 10^{11} M_{\odot}$ they find   $\rho_c=0.03 
M_{\odot}$/pc$^3$ in agreement with our value of
central density of $ 0.03 M_{\odot}$/pc$^3$. 

\section{Summary and conclusions}

We have calculated the evolution of the density profiles of 
self-interacting CDM haloes using a numerical approach based 
on the collisional Boltzmann equation. This study presents
two original points: ({\it i}) we were able to explore the 
halo density profiles covering about five orders of magnitude 
in the radius and for haloes from galaxy to galaxy cluster scales; 
({\it ii}) we used a self-interacting cross section inversely 
proportional to the collision velocity of the particles.
Analysing the results obtained with our code we conclude:
\begin{itemize}
\item [(1)] The interior of dark haloes may be deeply affected
   by self-interaction. A modest cross section value 
   ($\sigma v_{100} \approx 10^{-24}$ cm$^2$ GeV$^{-1}$) is enough
   to produce soft cores.
\item [(2)] Collisions between dark particles induce a thermalization
   process with the consequent appearance of a soft core. Thermal 
   equilibrium in the core is established with a few 
   collisions in a Hubble time ($\approx 4$).
\item [(3)] Soft cores appear already at the very early dynamical 
   evolution ($z \ge 40$); with time the central density decreases and
   the core size increases.    
\item [(4)] The observed scale invariance of the central density from galactic
   to galaxy cluster scales is
   reproduced only if the cross section is inversely proportional 
   to the collision velocity. 
\item [(5)] The core radius predicted by the model increases proportional
   to the halo maximum rotation velocity  
 in agreement with the 
   observed trend. 
\item [(6)] The observed central density of CL0024+1654 is consistent 
   with assuming a self-interacting dark matter cross section given by:
   $\sigma v_{100} \approx  10^{-24}$ cm$^2$ GeV$^{-1}$. 
\end{itemize}
The code is limited in analysing dark halo properties which depart
from spherical symmetry. Further studies concerning dark halo 
triaxiality and the evolution of substructures have been planned
using N-body techniques.

\section{Acknowledgments}
We are grateful to Vladimir Avila-Reese for a critical reading of an
early draft of the paper and to 
Carmelo Guzman, Liliana Hern\'andez and Gilberto 
Zavala for computing assistance. ED thanks Fondazione Cariplo for 
financial support and is grateful to
UNAM in Mexico City for its hospitality during the preparation
of this work.

\end{document}